\documentclass[preprint,preprintnumbers,amsmath,amssymb]{revtex4}
\usepackage{graphicx}
\usepackage{dcolumn}
\usepackage{bm}

\begin{document}
\title{Coupling between time series: a network view}
\author{S. Mehraban $^1$, A. H. Shirazi $^3$, M. Zamani $^2$, G. R. Jafari
$^{2,3}$ \\
{\small $^1$ Department of Physics, Sharif University of Technology, Tehran 11155-9161, Iran} \\
{\small $^2$ Department of Physics, Shahid Beheshti University,
G.C., Evin, Tehran 19839, Iran} \\
{\small $^3$ School of Nano-Science, Institute for Research in
Fundamental Sciences (IPM), Tehran, Iran}}
\date{\today}

\begin{abstract}

Recently, the visibility graph has been introduced as a novel view
for analyzing time series, which maps it to a complex network.
In this paper, we introduce a new algorithm of visibility,
"cross-visibility", which reveals the conjugation of two
coupled time series. The correspondence between the two
time series is mapped to a network, "the cross-visibility
graph", to demonstrate the correlation between them. We
applied the algorithm to several correlated and uncorrelated
time series, generated by the linear stationary ARFIMA
process. The results demonstrate that the cross-visibility
graph associated with correlated time series with power-law
auto-correlation is scale-free. If the time series
are uncorrelated, the degree distribution of their cross-visibility network
deviates from power-law. For more clarifying the process, we
applied the algorithm to real-world data from the financial
trades of two companies, and observed significant small-scale
 coupling in their dynamics.

\end{abstract}

\pacs{64.60.ae, 64.70.Tg}
\maketitle

\section{Introduction}

Coupled series, and series of long range
correlations could not be thoroughly understood, unless a
global view of them is represented~\cite{kantz1997nonlinear}. Complex networks~\cite{boccaletti2006complex, newman2003structure,
albert2002statistical}, on the other hand, have provided a
global understanding of multi-component systems. Recently, the
idea of complex networks has been implemented for single
variable systems~\cite{jafari2011coupled, shirazi2009mapping},
and mapping between time series and networks is proposed. For
example, nodes in such networks can be extracted by binning
the time axis or the level axis~\cite{vahabi2011analysis}.

While some features of time series are not simply observed with
standard approaches, they can be easily studied when converted
to a network~\cite{nunez2012visibility}. For example, financial trades, which their
dynamics seem to be alike to a white noise, include memory~\cite{liu1997correlations, pasquini1999multiscale, mantegna1999introduction} and
represent high clustering when converted to complex networks.
Moreover, investigating a map between time series and networks
is a method of constructing several prototypes of complex
networks~\cite{xie2011horizontal}.

According to a recent work~\cite{lacasa2008time}, the visibility algorithm has been introduced as a fast
computational algorithm of mapping a time series to a network.
Several properties of the time series are served in the
resulting network's structure. For example, a periodic series
results in a regular lattice, a fractal series results in a
scale-free network and a random series is in correspondence
with a random graph.

Visibility is the representation of the question that how much
local information is attained from a global structure. Based on this approach, new global properties are introduced for time series. For example, along with scale-free properties, small-world characteristics~\cite{strogatz2001exploring} are demonstrated for the fractional Brownian motion time series~\cite{lacasa2008time}. Also, in
order to distinguish randomness in time series, a modified
visibility algorithm~\cite{luque2009horizontal} is proposed.
Visibility algorithm is found to be an appropriate approach in
different interdisciplinary areas. For example, the method is
pertinent to social studies~\cite{fan2010fractal}. In
addition, mathematical structures such as fractal and complex
chaotic time series have been studied using the visibility algorithm~\cite{lacasa2008time, lacasa2010description}.

In this work, we introduce a new visibility algorithm, based
on the mutual information of two time series. We call this type
of visibility the cross-visibility. According to this method, the visibility criterion is represented for two time series, i.e., a time series looks at its components through the obstacles of a second time series. This approach is especially
appropriate for analyzing real-world data with power-law
autocorrelation. The algorithm provides a framework to
understand how complex time series lead each other, and determines
the direction of information flow between them. In addition,
coupling in different scales is demonstrated. Investigating the
cross-visibility between well-known correlated and uncorrelated
time series could be a criterion for characterizing the cross-visibility. For this purpose, we studied the fractional Gaussian noise
(fGn) time series.

Previously, it has been demonstrated that the Hurst exponent for the fractional Brownian motion series is distinguished by the structure of the degree distribution in their visibility graph~\cite{lacasa2009visibility}. We extended this analysis for correlated fGn time series, and characterized the power-law exponent in the resulting cross-visibility networks' degree distribution,
according to their Hurst exponent. Also, in order to depict the
implementation of the approach to real-world data, we used
the data from recent financial trades for two companies. We
depicted the information flow in several scales
between pertinent data.

\section{ The cross-visibility algorithm}

According to the standard visibility algorithm ~\cite{lacasa2008time}, every
component in the time series $\{y_i\}$, with $i=1 \ldots N$, is mapped to
a node of a graph. Node $i$ and node $j$ $(i<j)$ are connected, if they
are visible to each other without any obstacles between them:

\begin {equation}
i<\forall k <j; \hspace{1cm} y_k<y_i+\frac{y_j-y_i}{j-i}(k-i).
\end {equation}

If node $i$ is visible to node $j$, node $j$ will also visible
to node $i$. Therefore, the corresponding network is non-directional. Elements in the time series with amount
high above the others have high degrees, and create
the singular nodes of the network.

The cross-visibility algorithm maps two time series, $\{x_i\}$ and $\{y_i\}$, into two different networks, the cross-visibility networks. The algorithm consists of two major steps:

a. In order to make the time series comparable, they are normalized to reveal dimensionless variables. If the sequences are stationary, they are normalized to their mean and variances. Therefore, new
sequences are generated as $\{\hat{x_i}\}=\{\frac{x_i-\bar{x}}{\sigma_x}\}$ and
$\{\hat{y_i}\}=\{\frac{y_i-\bar{y}}{\sigma_y}\}$. $\bar {x}$
and $\bar{y}$ are the mean values of the series $\{x_i\}$ and
$\{y_i\}$, respectively; $\sigma_{x}$ and $\sigma_{y}$ are the
corresponding variances. If the series are non-stationary, they
are normalized to their corresponding maximum value.

b. Every component in the time series $\{\hat{x}_i\}$ is mapped to a node of a graph. Node $i$ is connected to node $j$, if:

\begin{equation}
i< \forall k <j; \hspace{1cm} \hat{y}_k \leq \hat{y}_i+\frac{\hat{x}_j-\hat{x}_i}{j-i}(k-i)
\label{TV}
\end{equation}
or
\begin{equation}
i< \forall k <j; \hspace{1cm} \hat{y}_k \geq \hat{y}_i+\frac{\hat{x}_j-\hat{x}_i}{j-i}(k-i).
\label{BV}
\end{equation}

This rule could be interpreted as node $i$ looking at the components of $\{\hat{x}_j\}$ time series,
through the obstacles of the shifted time series $\{\bar y_j\} = \{\hat{y}_j-\hat{y}_i+\hat{x}_i\}$. The structure of this network demonstrates how $\{x_i\}$ leads the time series $\{y_i\}$. Equation \ref{TV} refers to the visibility from the top view, and equation \ref{BV} demonstrates the visibility from the beneath view. We depict this graph as $G^{xy}$. However, another network, $G^{yx}$, could be generated by changing the role of $\{x_i\}$ and $\{y_i\}$.

The cross-visibility can bring a directional measure of
mutual information between the time series. If the time series $\{y_j\}$
is permuted randomly to construct a time series ( $\{y^p_j\}$) uncorrelated to
$\{x_j\}$, the structure of the cross-visibility
network $G^{x,y^p}$ deviates from the original one. Each network could be evaluated through its degree distribution.

\begin{figure*}
\vskip4mm
\includegraphics[width=17cm]{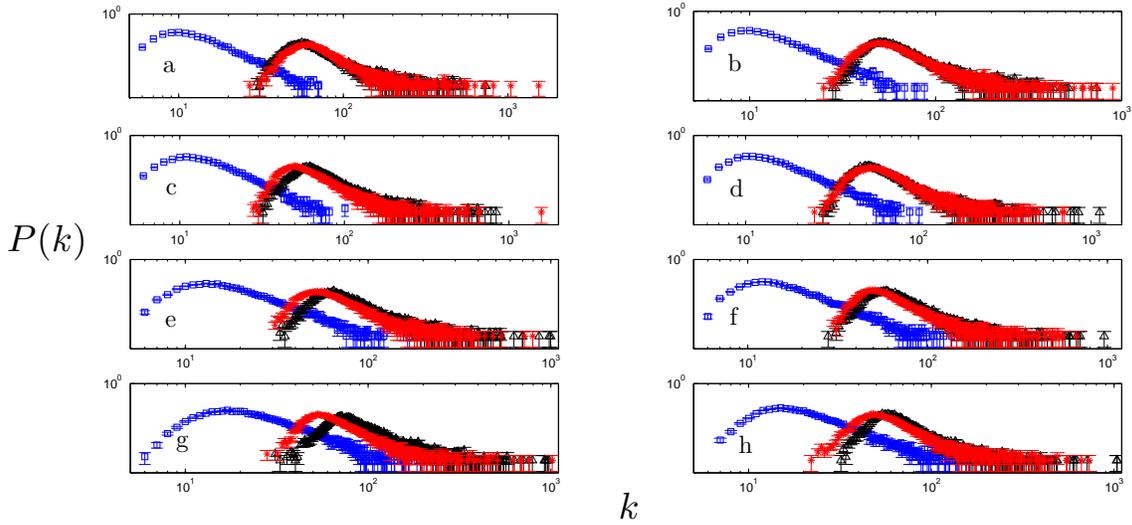}
\caption{(Color on-line) Degree distribution for several cross-visibility
networks corresponding to different Hurst exponents is depicted. (Blue) squares demonstrate the result
for the coupled series, (red) stars are for the uncoupled series,
and (black) triangles are for the permuted series. Coupling between several
series with $H\in\{0.6, 0.7, 0.8, 0.9, 1\}$ is investigated.
$0.6-0.7$ and $0.7-0.6$ are included in a and b, $0.6-0.8$ and
$0.8-0.6$ are included in c and d, $0.6-0.9$ and $0.9-0.6$ are
included in e and f, and $0.6-1$ and $1-0.6$ are included in g
and h, respectively. Both axes are sketched in logarithmic scale.}
\label{fig1}
\end{figure*}

\begin{figure*}
\vskip4mm
\includegraphics[width=15cm]{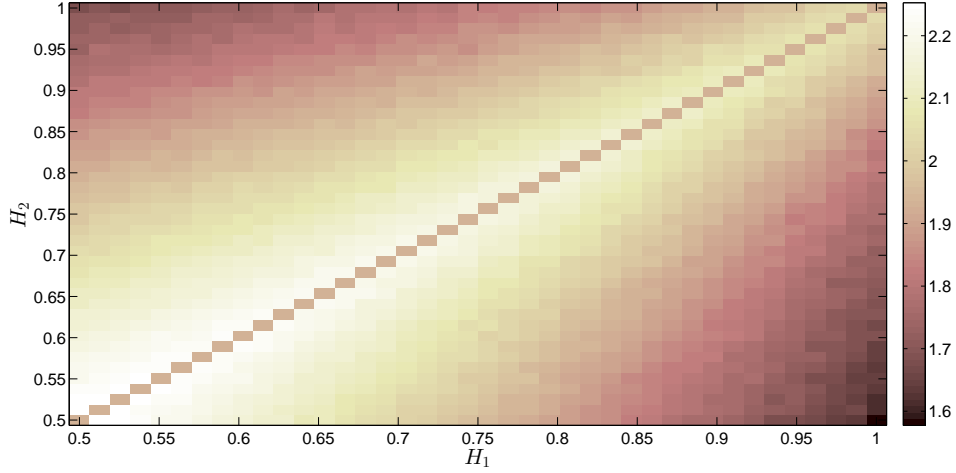}
\caption{(Color on-line) The color plot demonstrates the exponent of the power-law
degree distribution associated with the cross-visibility networks
of coupled series with $H\in[0.5,1[$. The result is sketched for the cross-visibility networks $G^{H_1, H_2}$, where $H_1$ and $H_2$ are the corresponding Hurst exponent for the first and the second time series, respectively.}
\label{fig2}
\end{figure*}

\begin{figure*}
\vskip4mm
\includegraphics[width=15cm]{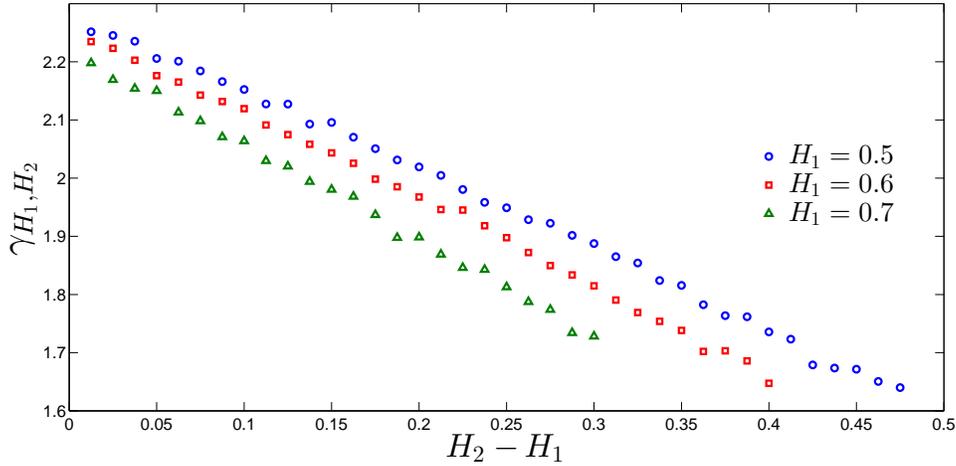}
\caption{(Color on-line) Power-law exponent for the degree distribution of cross-visibility graphs associated with coupled time series is sketched as a function of the difference between the corresponding Hurst exponents. The result is depicted for different Hurst exponents, $H_1$, for the cross-visibility network $G^{H_1,H_2}$. The power-law exponent decays linearly with $H_2-H_1$.}

\label{Hurst}
\end{figure*}
\begin{figure*}
\vskip4mm
\includegraphics[width=15cm]{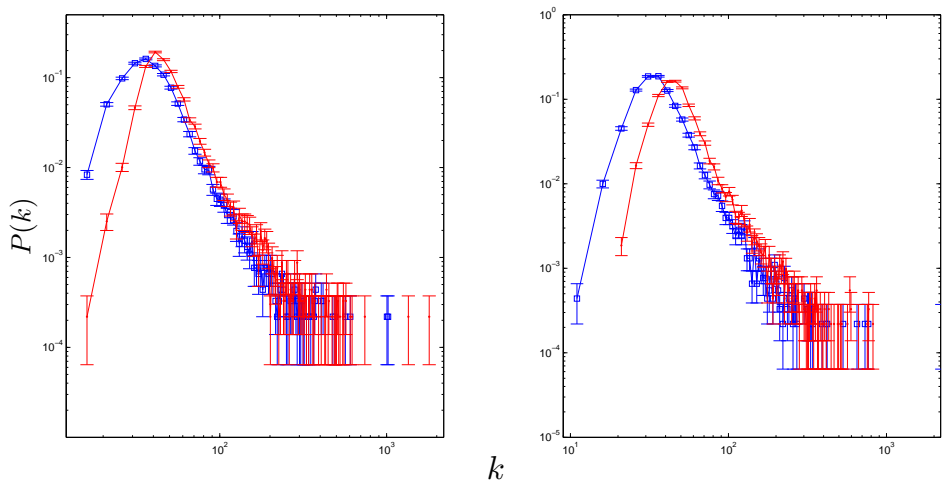}
\caption{(Color on-line) The closing price of simultaneously recorded
data for Microsoft and IBM is analyzed by the
cross-visibility algorithm. Degree distribution associated with
the original data is depicted by (blue) squares. (Red) dots depict the degree distribution for the cross-visibility network of permuted time series, $G^{x,y^p}$. The deviation is mostly concerned with the small scales, and also the pick of each distribution is shifted for uncoupled time series. Both axes are sketched in logarithmic scale.}
\label{fig3}
\end{figure*}
\begin{figure*}
\vskip4mm
\includegraphics[width=15cm]{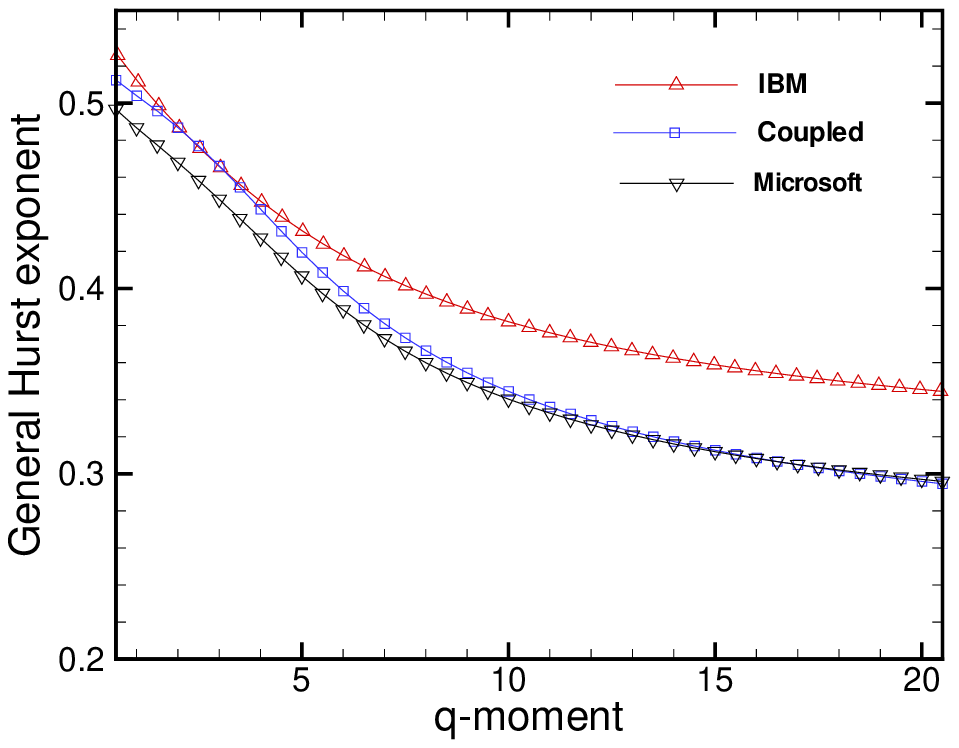}
\caption{(Color on-line) General Hurst exponents IBM, Microsoft and their
coupled estimated using multi-fractal detrended
cross-correlation analysis.} \label{fig4}
\end{figure*}
\subsection{Characterizing cross-visibility}

In order to interpret the results from the cross-visibility
algorithm, it is worthy to analyze the time series that are
confirmed to be coupled or uncoupled. We
investigated the time series generated by the linear stationary ARFIMA ~\cite {podobnik2007power, hosking1981fractional,granger1980long} process. The
analysis is a criteria for further investigations.

Through a stochastic recursive rule, the ARFIMA process generates a power law auto-correlated
series $\{y_k\}$:
\begin{equation}
y_i=\sum_{j=1}^{\infty}a_j(\rho)y_{i-j}+\eta_i,
\label{ARFIMA}
\end{equation}
where, $a_j(.)$ is a weight, defined due to an independent real
variable $0<\rho<\frac{1}{2}$ as
$a_j(\rho)=\frac{\Gamma(j-\rho)}{\Gamma(-\rho)\Gamma(1+j)}$.
$\eta_i$, the error term, is a member of an i.i.d sequence, which consists of random normal distribution samples.
The Hurst exponent, $H$, is related to $\rho$ as $H=0.5 +
\rho$ ~\cite{hosking1981fractional}. These time series depict power-law autocorrelation~\cite{horvatic2011detrended}, and If exactly same error sequence is used in generating the
two time series, they show power-law cross correlation. In contrast, independent
error terms reveal uncoupled time series.

We applied the cross-visibility algorithm to coupled
time series, with different independent terms $\rho\in
[0, 0.5[$, which are in correspondence with Hurst exponents $H
\in [0.5,1[$. FIG. \ref{fig1} depicts the degree distribution
associated with each network. A power-law degree distribution
is demonstrated for the coupled series. The cross-visibility
network for the uncoupled time series have completely
different structure in contrast to the corresponding network of
coupled series (stars and triangles in FIG. \ref{fig1}). Moreover, the average degree of the distributions is higher for the uncoupled series.

If the time series are correlated, the corresponding exponent of the power-law
degree distributions, depicted as $\gamma_{xy}$ for $G^{xy}$, determine how the series are leading each other.
$\gamma_{xy}$ for several series with Hurst exponents $H \in [0.5, 1[$ is depicted as a color plot in
FIG. {\ref{fig2}}. In order to determine the exponent of the degree distribution we used a maximum likelihood estimation (MLE) ~\cite{newman2005power}. In FIG. \ref{Hurst} a linear dependency between $\gamma_{H_1,H_2}$ and $H_2-H_1$ is depicted. This result demonstrates that the Hurst exponents could be distinguished by evaluating the power-law exponents. The result is sketched for several Hurst exponents $H_1$.

\subsection{Application in empirical data}

For real-world data, it is important to confirm if the
time series are coupled and to what extent they follow each
other. In order to determine the coupling between the two
time series, $\{x_i\}$ and $\{y_i\}$, we put the
associated degree distribution of their cross-visibility networks into contrast with the corresponding network of permuted time series.

We analyzed the cross-visibility between the closing price of two
companies, Microsoft and IBM. The cross-visibility method is
applied to the normalized logarithm of the prices. For each
company, return prices are stationary and are normalized to their average and
standard deviations. The degree distribution of
the cross-visibility graphs associated with original series and
permuted time series is depicted in Fig. \ref{fig3}. For the original series,
the associated degree distribution is a power-law. However, if
one of the series (the second series) is permuted, the
associated degree distribution deviates from the original
distribution. The most significant deviations are concerned
with the small degrees. This however means that there is strong
correlation in small fluctuations. Deviation in the tail of the
distributions corresponding to high degrees is negligible.
These time series do not follow each other in their high amplitude
fluctuations. The value of the degree corresponding to the pick
of the distribution is a characteristic of the lag time response between
the two time series.

It is worthwhile to compare the results with the systematic obtaining of the cross-correlation analysis (DXA)
\cite{podobnik2008detrended,zhou2008multifractal,shadkhoo2009multifractal}. Fig. \ref{fig4} shows the results for DXA, and indicates how Microsoft leads coupling in higher moments and
IBM in lower moments. One important difference is that DXA is
a symmetric method to analyze the cross-correlation, but the cross-visibility introduces two networks of coupling
between the two time series.

\section{Conclusions}

- Understanding the coupling between two time series and how
information transfers between them is a crucial concept in
various areas of complex systems. For this purpose, in this
paper we depicted this coupling as a network map.

- In general, as the coupling between two time series is
asymmetric, there are two networks to present the coupling
between them.

- Our results demonstrate that in average, the number of
degrees corresponding to the coupled series is less than that
of the uncoupled data. There is a shift in $P(k)$ to higher
degrees in uncoupled time series, which could be a criteria to
measure the value of coupling.

- Our findings show that, increasing the difference between the
Hurst exponent of two fGn time series, the corresponding degree
distribution decays with slower slope in higher degrees,
which means higher coupling or higher correspondence between the two
time series. This approach is appropriate for investigating coupling
for different empirical data. In addition, through this
approach the conjugation in different scales between
time series could be studied. In the end, using this approach, the coupling between
Microsoft and IBM has been viewed in two coupled networks.

\bibliography{References}

\end{document}